\pdfoutput=1

\documentclass[twocolumn,           
               showpacs,            
               preprintnumbers,     
               aps,                 
               prd,                 
               a4paper,             
               superscriptaddress,  
               nofootinbib,         
               tightenlines,        
               floats,floatfix      
               ]{revtex4}

\usepackage{graphicx}
\usepackage{subfigure}
\usepackage{latexsym}
\usepackage{amsmath,amssymb}        
\usepackage[draft=false]{hyperref}

\input colordvi

\begin{document}



\title{Triple unification of inflation, dark matter, and dark energy
  using a single field} 
\author{Andrew R.~Liddle}
\affiliation{Astronomy Centre, University of Sussex, Brighton BN1 9QH,
United Kingdom}
\author{C\'edric Pahud}
\affiliation{Astronomy Centre, University of Sussex, Brighton BN1 9QH,
United Kingdom}
\affiliation{ Institut d'Astrophysique de Paris, UMR 7095-CNRS, Universit\'e
Pierre et Marie Curie, 98bis boulevard 
Arago, 75014 Paris, France}
\author{L. Arturo Ure\~na-L\'opez}
\affiliation{Instituto de F\'isica de la Universidad de Guanajuato,
C.P.~37150, Le\'on, Guanajuato, M\'exico}  
\date{\today}


\begin{abstract}
We construct an explicit scenario whereby the same material driving
inflation in the early Universe can comprise dark matter in the
present Universe, using a simple quadratic potential. Following
inflation and preheating, the density of inflaton/dark matter
particles is reduced to the observed level by a period of thermal
inflation, of a duration already invoked in the literature for
other reasons. Within the context of the string landscape, one can
further argue for a non-zero vacuum energy of this field, thus
unifying inflation, dark matter and dark energy into a single
fundamental field.
\end{abstract}

\maketitle


\section{Introduction}

In a recent paper \cite{LU}, two of us proposed a general scenario for
unification of dark matter and inflation into a single field. The key
ingredient is the survival of a residual amount of the inflaton
field's energy density, which undergoes coherent oscillations and can
serve as a cold dark matter candidate. In the context of the string
landscape, one can further argue for a non-zero vacuum energy of this
field on anthropic grounds, thus providing a single description of the
three key unknowns of modern cosmology, namely dark energy, dark
matter, and the material responsible for early Universe inflation.

In practice, however, realizing this scenario is non-trivial, due to
the need for a long radiation-dominated era of the Universe
encompassing the nucleosynthesis period. This requires that the
amplitude of scalar field oscillations be extremely small after the
energy trapped in the inflaton is released into normal
particles. Preheating scenarios can provide part of the required
reduction of the oscillation amplitude, but still leave it too high
and in conflict with the observed dark matter to radiation density
ratio.\footnote{This holds for the original four-legs interaction
studied in the preheating
literature \cite{Kofman:1994rk,Shtanov:1994ce,Kofman:1997yn}, though a
complete decay of the inflaton can be obtained from the introduction
of other couplings~\cite{Dufaux:2006ee}.}

In this paper, we explore a modification to the original scenario of
Ref.~\cite{LU}. As we shall discuss below, after preheating the
Universe undergoes a short period of radiation quickly followed by a
period of matter domination driven by the relic energy density of the
inflaton field itself. This early matter domination period is
interrupted by a short second period of inflation, known as thermal
inflation, driven by a separate field and perhaps associated with the
supersymmetry breaking transition. We find that thermal inflation can
reduce the oscillation amplitude of the scalar field to the desired
level, and then provide a proper reheating of the Universe.

\section{Cosmological evolution}

For definiteness, we consider throughout the model of Ref.~\cite{LU}
where the inflaton $\phi$ has potential $V_0 +
\frac{1}{2} m^2 \phi^2$. Here $V_0$ has the small value needed to
explain the observed dark energy density, and otherwise does not play
a significant role. For sufficiently large $|\phi| \gtrsim m_{\rm
Pl}$, this potential drives inflation and produces density
perturbations in agreement with observations provided $m \simeq
10^{-6} m_{\rm Pl}$. Subsequently $H \ll m$ at all times, where $H$ is
the Hubble parameter, and the $\phi$ field oscillates rapidly on the
Hubble timescale. Such an oscillating field behaves as cold dark
matter, both in the redshifting of the mean density $\rho_\phi \propto
a^{-3}$ and in the evolution of perturbations.

Unless some mechanism exists to reduce the energy density of the
oscillating field, and indeed to transform some of it into
conventional material, it is not possible to recover a satisfactory
Big Bang cosmology. The original resolution was reheating --- the
complete transfer of energy from the inflaton via single-particle
decays. Later coherent decays, known as preheating
\cite{Kofman:1994rk,Shtanov:1994ce,Kofman:1997yn,Bassett:2005xm}, were
invoked as well. Such decays may be extremely efficient when the
inflaton oscillations are large, but if the only interactions present
are annihilations, the process will necessarily shut off once the
density reduces. This led Kofman, Linde, and
Starobinsky \cite{Kofman:1994rk,Kofman:1997yn} to propose that the
residual field could act as dark matter, but in fact detailed
calculations \cite{LU} show the relic abundance is far too high under
standard assumptions. It has usually thus been considered that
preheating is followed by a period of reheating leading to complete
decay of the inflaton field.

Having recognized that an inefficient reheating is a main concern in
our unification scenario, the authors in
Refs.~\cite{Cardenas:2007xh,Panotopoulos:2007ri}\footnote{There
exists a sneutrino (which is a scalar field) unification model for
inflation and dark matter \cite{Panotopoulos:2007ri,Lee:2007mt}, with
similar properties to our phenomenological model; under certain
conditions, our approach also applies to it.} suggest that plasma mass
effects \cite{Kolb:2003ke} could provide the mechanism for an
incomplete reheating after inflation. The idea is that the decay of
the inflaton field is kinematically forbidden in part of the reheating
phase. However, the inflaton field is free to decay once it becomes
subdominant with respect to the radiation fluid (see the paragraph
after Eq.~(9) in Ref.~\cite{Kolb:2003ke}), so thermal masses cannot be
thought of by itself as a mechanism for incomplete reheating.

We can consider three main possibilities for reducing this excess
abundance, while leaving a relic level of oscillations capable of
acting as cold dark matter. The first is to modify the shape of the
inflaton potential. However it is easy to show that the required level
of post-inflationary oscillations is too small for such a modification
to work; inflation must end long before the field is near enough the
minimum to give the right abundance. This approach is therefore
fruitless. The second possibility is to modify the reheating process
so that it leaves a relic abundance level; this was the approach of
Ref.~\cite{LU}, who chose a phenomenological form for the decay rate
intended to correspond to particles which only had annihilation routes
rather than decay routes, thus permitting incomplete reduction of the
inflaton oscillations. However fine-tuning of the decay rate,
unmotivated by fundamental theory, is required to make this scenario
work.

In this paper we consider a third possibility, which appears more
attractive and natural, which is to consider a brief period of
inflation at lower energy densities. Such a period, often called
\emph{thermal inflation}
\cite{Lyth:1995hj,Barreiro:1996dx,Lyth:1995ka}, was introduced in
order to remove possible relic abundance problems left over by the
original high-energy inflation period. This second period would be too
short to imprint any new large-scale perturbations, but would reduce
the abundance of any relic particles compared to the ultimate
radiation background. An oscillating scalar field would have its
density reduced by this mechanism.

\begin{figure}[t]
\centering
\includegraphics[width=8.7cm]{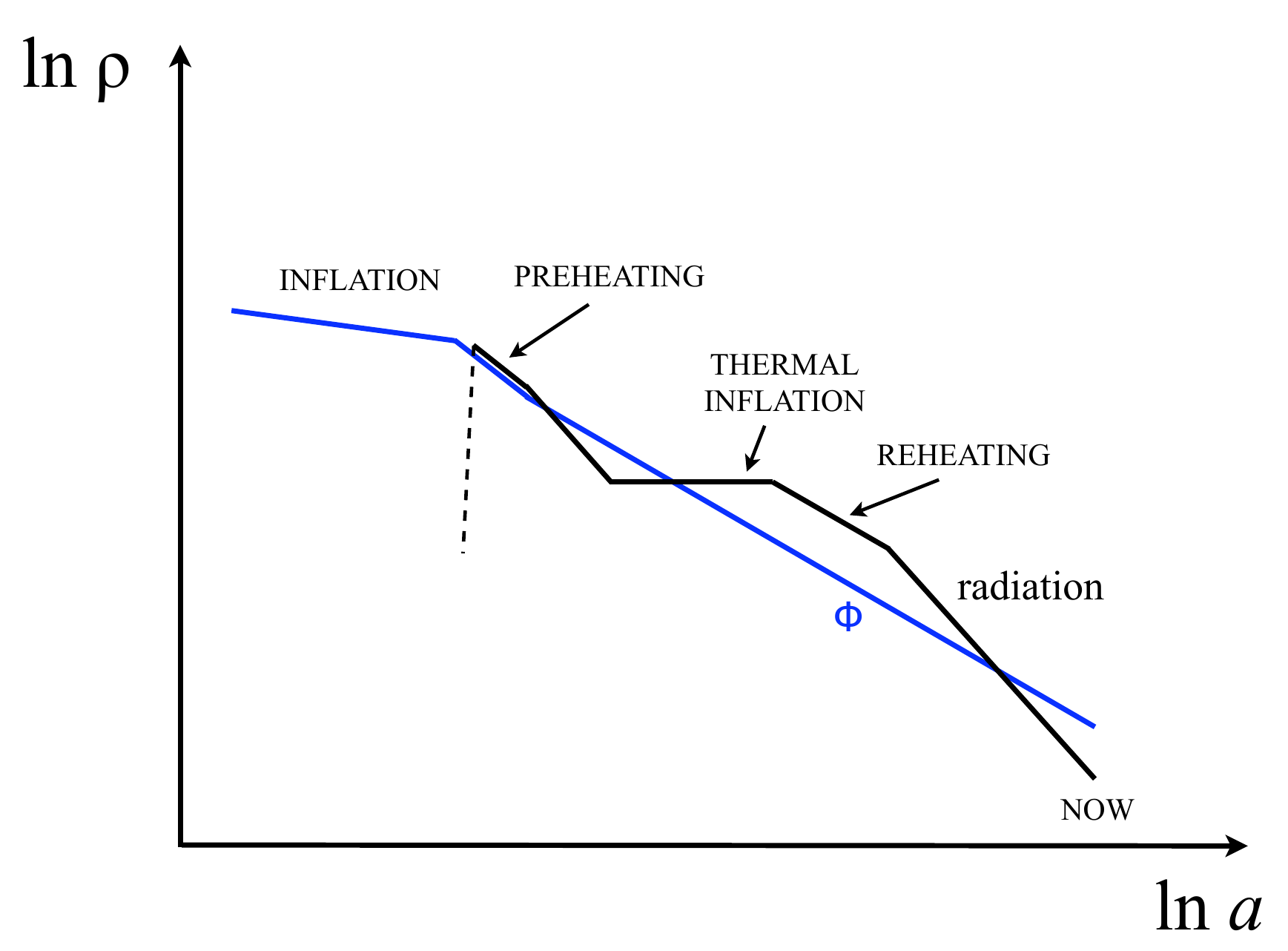}
\caption{\label{f:densities} A schematic of the evolution of the
densities throughout the Universe's evolution. One (blue) line shows
the density of the $\phi$ field, and the other (black) line the
combined density of all other materials. The latter changes shape
depending whether this combined density is dominated by radiation or
by the material driving thermal inflation. We also include a reheating
period after thermal inflation. [To avoid confusing the diagram, we do
not show the emergence of non-relativistic baryons from the
relativistic fluid at late times.]}
\end{figure}

For future reference, Fig.~\ref{f:densities} shows a schematic of
the Universe's evolution for our proposed scenario.  As we will show,
the required reduction, assuming a period of preheating after
inflation but no reheating, is achieved provided thermal inflation
lasts for around 12 $e$-foldings. This is in agreement with
  the duration already suggested in the
  literature \cite{Lyth:1995hj,Barreiro:1996dx}.

\section{A detailed scenario}

We first revisit the calculation of the dark matter mass-per-photon
for our scalar field, which ultimately gives the strongest constraint
on the parameters of our model.

Let us denote by $t_\ast$ the time after which the required Hot Big
Bang (HBB) cosmology is recovered\footnote{Notice that the meaning of
$t_\ast$ is changed with respect to Ref.~\cite{LU}, where it was
intended to denote the time at which the equality $H=m$ was
achieved.}; hence, the averaged scalar field energy density will be
given by $\rho_\phi = m^2 \phi^2_\ast a_\ast^3 /a^3$ for $t >
t_\ast$. Hereafter, all quantities with an asterisk denote values at
time $t_\ast$.

As in Ref.~\cite{LU}, we define the scalar field dark
matter-mass-per-photon as $\xi_{{\rm dm}} \equiv \rho_\phi/n_\gamma$,
and we assume expansion at constant entropy implying that $\xi_{\rm
  dm}/g_{\rm S}$ remains constant where $g_{\rm S}$ is the entropic
degrees of freedom, usually very similar to the relativistic degrees
of freedom that we denote here by $g_{\rm E}$ \cite{Kolb:1990vq}. It
is straightforward to show that, for any time $t > t_\ast$,
\begin{equation}
  \frac{\xi_{\rm dm}}{m_{\rm Pl}} = \frac{\pi^2}{2 \zeta(3)}
  \frac{g_{\rm S}(T)}{g_{\rm S}(T_\ast)} \frac{m^2}{m^2_{\rm Pl}}
  \frac{\phi^2_\ast}{m^2_{\rm Pl}} \frac{m^3_{\rm Pl}}{T^3_\ast} \,
  , \label{eq:phi-dm}
\end{equation}
where $T$ is the temperature of the Universe, as measured from the
relativistic particles in thermal equilibrium at time $t$.

Eq.~(\ref{eq:phi-dm}) contains two free parameters, which are the
scalar field $\phi_\ast$ and the temperature $T_\ast$ at the beginning
of the HBB; equivalently, we shall call this the time at the end of
reheating. It is then necessary to predict the aforementioned values
and determine whether they can match the observed value of $\xi_{\rm dm}$.

In the early Universe, there is first a stage of slow-roll inflation,
at the end of which the inflaton field value is $\phi_{\rm end}\simeq
0.28 m_{\rm Pl}$. Then a preheating stage starts in which part of the
inflaton energy density is converted into relativistic degrees of
freedom. We assume the simplest model of preheating \cite{Bassett:2005xm}, 
in which the inflaton field is coupled to a massless scalar field
$\chi$ through the four-legs interaction term 
\begin{equation}
  V_{\rm int} = \frac{g^2}{2} \phi^2 \chi^2 \, , \label{eq:4legs}
\end{equation}
where $10^{-10} < g^2 < 10^{-5}$ is the typically-considered range for
the coupling constant.  

The preheating process ends once the inflaton amplitude is of the
order $\phi_{\rm pr} \simeq m/g$, at which point the ratio between
relativistic ($\chi$ and $\phi$ quanta) and non-relativistic degrees
of freedom (coherent oscillations of $\phi$) $\rho_r /\rho_\phi$ is of
order of a few \cite{Podolsky:2005bw}. In such a case, we cannot
expect a prolonged radiation-dominated era after preheating; rather, we
expect the appearance of a matter-dominated era just a few $e$-folds
after the end of preheating when the coherent $\phi$ field comes back
into domination.

Alternative coupling terms in the potential, such as three-leg decay
interactions, can lead to a complete decay of the inflaton field
\cite{Dufaux:2006ee}; that would also happen in cases where the
inflaton field is coupled to fermionic fields
\cite{Kofman:1997yn}. However, we do not allow such couplings for the
inflaton field in our model, for instance by presuming that the $Z_2$
symmetry $\phi \leftrightarrow -\phi$ is (almost) exact. The $\phi$
field therefore survives right through to the present; however if the
radiation simply redshifts away as normal its density will be far too
low relative to that of $\phi$ by the present.

Instead, our proposal is that the subsequent evolution of the Universe
raises the radiation energy density back above that of the $\phi$
field, so as to re-establish a standard Hot Big Bang evolution.  After
the preheating process, the energy density of the Universe is composed
of relativistic particles and non-relativistic matter represented by
the coherent oscillations of the inflaton field. We now assume that
there is a second scalar field, initially part of the relativistic
thermal bath, that will drive thermal inflation in a later stage. This
second field, known as the \emph{flaton} field, has thermal
corrections to its potential which trap it in a false vacuum with
energy density denoted by $\hat{V}$. A hat will be used henceforth to
denote quantities related to the flaton field.

According to the standard picture of thermal inflation
\cite{Lyth:1995hj,Barreiro:1996dx,Lyth:1995ka}, an inflationary stage
starts once the false vacuum energy dominates over the radiation
fluid; this happens once the temperature of the latter is $T <
\hat{V}^{1/4}$. Thermal inflation then ends once the thermal
corrections to the potential are insufficient to oppose the underlying
symmetry-breaking (SB) potential, so that the thermal inflaton can
escape from its false vacuum and undergoes an SB transition. This
happens once the temperature of the Universe is below the flaton mass
scale, $T < \hat{m}$.

In our scenario the sequence is a little different, as seen in
Fig.~\ref{f:densities}, because the flaton density is initially
subdominant to the oscillating $\phi$ field. However after some
interval the $\phi$ density falls below it and thermal inflation
starts; sometime afterwards the SB transition then takes
place.

After the preheating process the inflaton field
redshifts as cold dark matter, $\phi \propto a^{-3/2}$, and we can
calculate the total dilution of the inflaton field from the
end of preheating up to the SB process. The square of the inflaton
field at the end of thermal inflation is given by
\begin{equation}
  \phi^2_{\rm SB} = \phi^2_{\rm pr} \left( \frac{a_{\rm pr}}{a_{\rm
      SB}} \right)^3 = \phi^2_{\rm pr} \frac{g_{\rm S}(T_{\rm
      SB})}{g_{\rm S}(T_{\rm pr})} \left( \frac{\hat{m}}{T_{\rm pr}}
  \right)^3 \, . \label{eq:phi-SB0}
\end{equation}
To obtain the above equation we are assuming both entropy conservation
and that the radiation fluid is in thermal equilibrium. $T_{\rm pr}$
and $T_{\rm SB}=\hat{m}$ are the values of the temperature at the end
of the preheating stage and at the SB process, respectively; likewise,
$a_{\rm pr}$ and $a_{\rm SB}$ are the corresponding values of the scale
factor.

Once thermal equilibrium is attained at the end of preheating
\cite{Bassett:2005xm,Podolsky:2005bw}, the usual formula for the
temperature of the radiation fluid applies, $\rho_{r,{\rm pr}} =
(\pi^2/30) g_{\rm E}(T_{\rm pr}) T^4_{\rm pr}$. Recalling that
$\rho_{r,{\rm pr}} \simeq \rho_{\phi,{\rm pr}}$, then $T_{\rm pr}
\simeq (30/\pi^2)^{1/4} g^{-1/2} g^{-1/4}_{\rm E}(T_{\rm pr}) \,
m$.\footnote{Incidentally, thermal inflation can resolve the relic
abundance troubles, e.g.\ the gravitino, that such a high temperature
$T_{\rm pr} \sim m \simeq 10^{13}{\rm GeV}$ may lead to \cite{gravitino}.}
Thus, from Eq.~(\ref{eq:phi-SB0}), the total dilution of the inflaton
field is largely determined by the mass scales of the inflationary
fields,
\begin{equation}
  \phi^2_{\rm SB}  \simeq \frac{\pi^{3/2}}{30^{3/4}}
  \frac{g_{\rm S}(T_{\rm SB})}{g_{\rm S}(T_{\rm pr})}
  \frac{g^{3/4}_{\rm E}(T_{\rm pr})}{g^{1/2}} \frac{\hat{m}^3}{m^3} \,
  m^2\, . \label{eq:phi-SB} 
\end{equation}

The last process is the reheating of the Universe at the end of
thermal inflation. We shall assume that each flaton particle decays at
a single-particle decay rate $\Gamma$, which is a new free parameter
in our phenomenological approach. In principle the value of $\Gamma$
can be estimated in terms of $\hat{m}$ and $\hat{V}$
\cite{Barreiro:1996dx}, as we discuss later.

The Universe is reheated when $\Gamma \simeq H_\ast$, where $H_\ast$
is the Hubble rate at the beginning of the HBB. In between, the
Universe is dominated by the energy density of the oscillating flaton
field (which redshifts as $a^{-3}$), so that the change in the scale
factor is given by
\begin{equation}
  \frac{a^3_{\rm SB}}{a^3_\ast} \simeq 
  \frac{H^2_\ast}{H^2_{\rm SB}} \simeq
  \frac{3m^2_{\rm Pl}\Gamma^2}{8\pi \hat{V}} \, . \label{eq:SB-ast}
\end{equation}
The inflaton field is further affected by this expansion as well, so
that we get
\begin{equation}
  \frac{\phi^2_\ast}{m^2_{\rm Pl}} = \frac{\phi^2_{\rm SB}}{m^2_{\rm
      Pl}} \frac{a^3_{\rm SB}}{a^3_\ast} \simeq
  \phi^2_{\rm SB} \frac{3 \Gamma^2}{8\pi \hat{V}} \,
  , \label{eq:phiSB-ast}
\end{equation}
where $\phi^2_{\rm SB}$ is given in Eq.~(\ref{eq:phi-SB}). Finally,
the reheating temperature $T_\ast$ is estimated to be
\cite{Kofman:1997yn}
\begin{equation}
  T_\ast = \left( 90/ 8 \pi^3\right)^{1/4} g^{-1/4}_{\rm E}(T_\ast)
  \sqrt{m_{\rm Pl}\Gamma} \, . \label{eq:temp-ast}
\end{equation}

We are now in a position to use the dark matter constraint from
Eq.~(\ref{eq:phi-dm}), which now takes the form
\begin{eqnarray}
\frac{\xi_{\rm dm}}{m_{\rm Pl}} &\simeq&
\frac{3\pi}{16 \zeta(3)}
\frac{g_{\rm S}(T_{\rm SB})}{g_{\rm S}(T_{\rm pr})}
\frac{g_{\rm S}(T)}{g_{\rm S}(T_\ast)}
g^{3/4}_{\rm E} (T_\ast) g^{3/4}_{\rm E} (T_{\rm pr}) \nonumber \\
&& \times \, \left( \frac{3}{8\pi} \right)^{1/4} g^{-1/2}
\frac{m}{\hat{m}} \left( \frac{\hat{m}}{\hat{V}^{1/4}} \right)^4
\sqrt{\frac{\Gamma}{m_{\rm Pl}}}\, . \label{eq:phi-dm-1}
\end{eqnarray}

The measured value of the current dark matter mass per photon is
$\xi_{{\rm dm},0}=2.4 \times 10^{-28} m_{{\rm Pl}}$ using values from
WMAP5 \cite{Dunkley2008}. We shall take that $g_{\rm E}(T) \simeq
g_{\rm S}(T) \simeq 100$ for temperatures $T \geq T_\ast$, and
\mbox{$g_{\rm S} (T_0) = 3.9$}, where `$0$' indicates present values;
Eq.~(\ref{eq:phi-dm-1}) then becomes
\begin{equation}
  g^{-1/2} \, \frac{m}{\hat{m}} \left( \frac{\hat{m}}{\hat{V}^{1/4}}
  \right)^4 \sqrt{\frac{\Gamma}{m_{\rm Pl}}} \simeq 1.4 \times
  10^{-29} \, . \label{eq:main}  
\end{equation}

We define the number of $e$-folds of thermal inflation as $N_{\rm
TI}\equiv \ln (\hat{V}^{1/4}/\hat{m})$, whereas we denote the number
of $e$-folds between the end of thermal inflation and the completion
of reheating, from Eq.~(\ref{eq:SB-ast}), as
\begin{equation}
  N_{\rm reh} \equiv \frac{1}{3} \ln \frac{8\pi
    \hat{V}}{3 m^2_{\rm Pl}\Gamma^2} \, .
\end{equation}
Thus, an equivalent expression for Eq.~(\ref{eq:main}) is, in terms of
the above-defined $e$-folding numbers,
\begin{equation}
  N_{\rm TI} + \frac{1}{4} N_{\rm reh} \simeq 18 - \ln
  \, g^{1/6} \, , \label{eq:main-1}
\end{equation}
where we have used $m/m_{\rm Pl} \simeq 10^{-6}$. For the expected
range $10^{-10} < g^2 < 10^{-5}$, the last term on the right-hand side
contributes one to two extra $e$-folds.

Equation~(\ref{eq:main-1}) is our main result, giving the duration of
thermal inflation and subsequent reheating required to give a viable
Universal history.  We now investigate how achievable it is.  The only
genuinely free parameter of our model is the decay width $\Gamma$,
which determines $N_{\rm reh}$ (the dependence on $g$ over its
expected range is modest). The reason is that thermal inflation
parameters are expected to lie in more or less definite ranges of
energy \cite{Lyth:1995hj}. The mass of the flaton field should be of
the order of $\hat{m} \simeq 10^2$ to $10^3$ GeV, and on general
grounds we expect $\hat{V}^{1/4} \simeq 10^7$ to $10^8$ GeV, so that
$N_{\rm TI} \simeq 11$ with an uncertainty of one or two in either
direction. This could be increased by having more than one period of
thermal inflation, but we do not need this.

\begin{figure}[t]
\centering
\includegraphics[width=8cm]{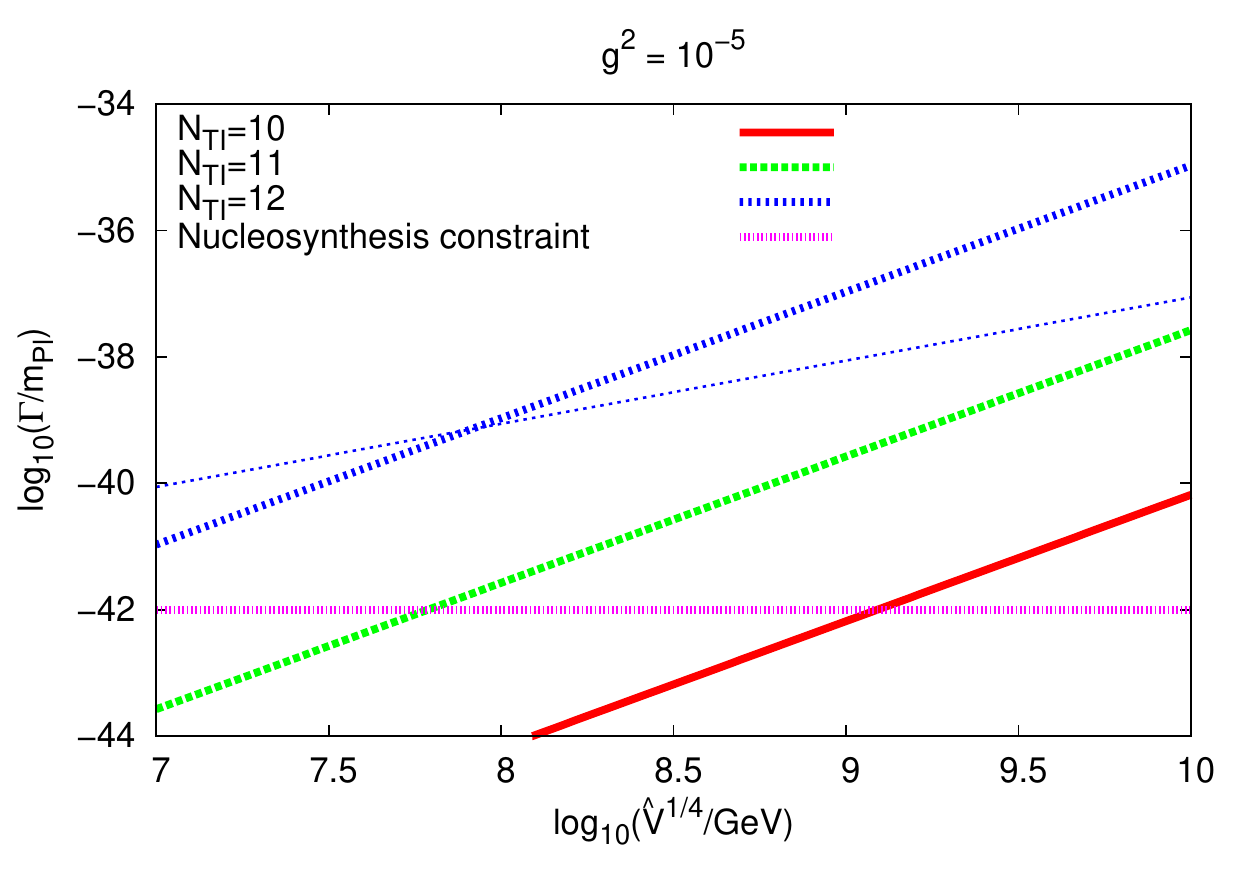}\\
\includegraphics[width=8cm]{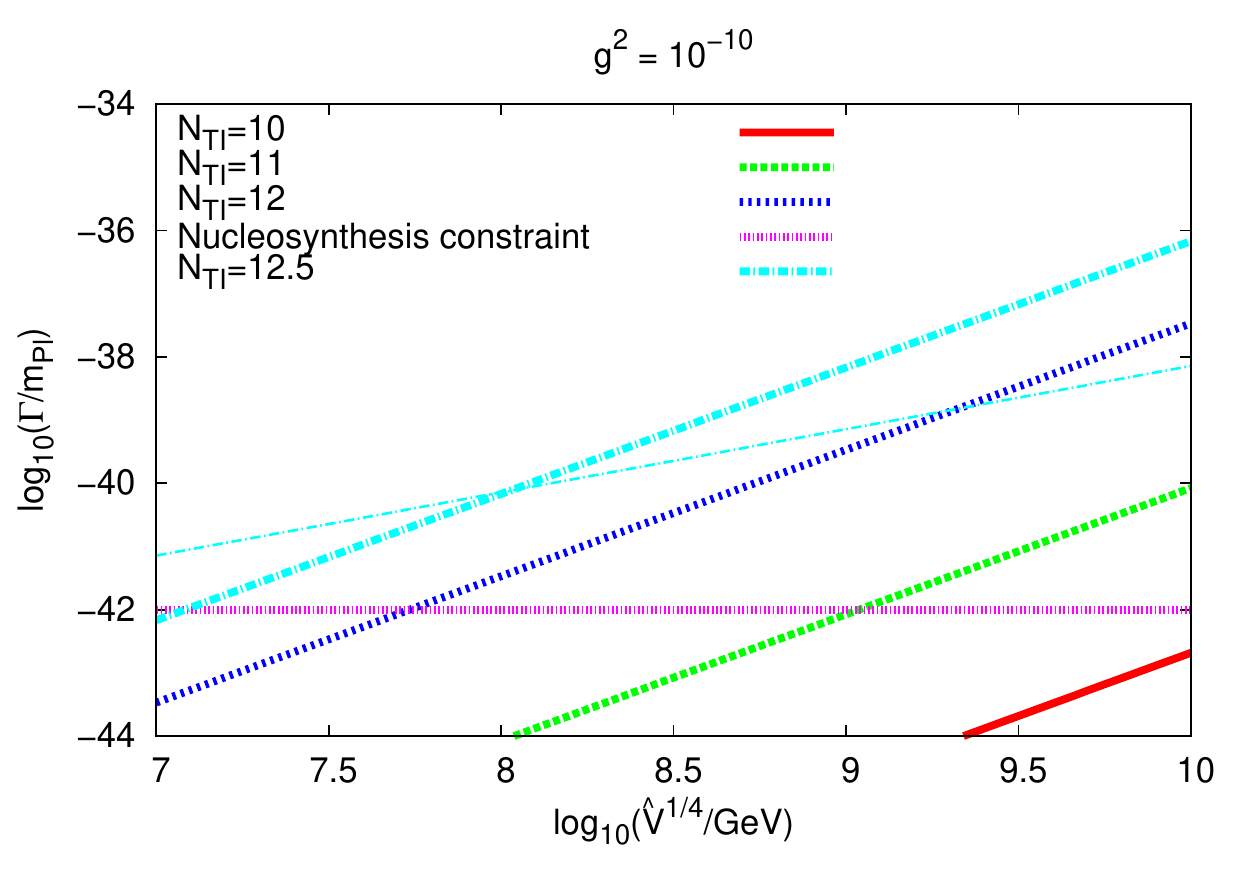}\\
\caption{\label{f:fig2} The lines in both figures show the value of
  $\Gamma$ required to satisfy the abundance constraint
  Eq.~(\ref{eq:main-1}), as a function of $\hat{V}$, for some fixed
  values of $N_{\rm TI}$ (note that the value of $N_{\rm TI}$ itself
  depends on both $\hat{V}$ and $\hat{m}$) and the preheating coupling
  term $g^2$. Only models with large enough values of $N_{\rm TI}$ lie
  above the nucleosynthesis constraint line $\Gamma > 10^{-42}m_{\rm
    Pl}$. We also show the thermal inflation estimation of the decay width
  $\Gamma$ (thin lines) for the cases $N_{\rm TI}= 12$ (top) and
  $N_{\rm TI}= 12.5$ (bottom); see text for details. As read from the crossing of the
  corresponding lines, models with flaton parameters of the order of
  $\hat{m} \simeq 10^3 \, {\rm GeV}$ and $\hat{V}^{1/4} \simeq 10^8 \,
  {\rm GeV}$ are able to satisfy all constraints.}
\end{figure}

The decay width is sandwiched by two limits: that the
decay should take place after thermal inflation is complete, $\Gamma <
H_{\rm SB} \simeq 10^{-24} m_{\rm Pl}$, and that it should be complete
before the run-up to nucleosynthesis begins at around 10 MeV,
requiring $\Gamma > 10^{-42} m_{\rm Pl}$.
Figure~\ref{f:fig2} shows the required value of $\Gamma$ to satisfy
the observational constraint~(\ref{eq:main-1}), as a function of
$\hat{V}$ and for some different values of $N_{\rm TI}$. We see that
the nucleosynthesis constraint can readily be satisfied provided
$N_{\rm TI}$ and $\hat{V}$ are large enough, and that suitable values
lie well within the expected range.

Actually, one can arguably justify that the typical decay
width of flaton particles is of the form $\Gamma \simeq 10^{-2}
\hat{m}^5 /\hat{V}$ \cite{Barreiro:1996dx}. If we plot this in
combination with Eq.~(\ref{eq:main-1}) in Fig.~\ref{f:fig2}, we find that the favoured flaton
parameters are $\hat{m} \simeq 10^3 \, {\rm GeV}$ and
$\hat{V}^{1/4} \simeq 10^8 \, {\rm GeV}$.

We therefore conclude that thermal inflation, with properties already
well established in the literature, can indeed dilute the
inflaton density sufficiently that it can act as dark matter.

\section{Conclusions and discussion}

The task of arranging that a residual inflaton density survives
to act as dark matter is a challenging one, but unification of two
normally unconnected sectors of cosmological modeling would be a
valuable reward (within the string landscape picture we can even argue that the same potential also gives rise to dark energy \cite{LU}). In this paper, we have shown that one option to
achieve this is to exploit the uncertainty in cosmological dynamics
during the long period from the end of inflation up to
nucleosynthesis. In particular, we have found that a period of thermal
inflation during this epoch has exactly the desired effect, reducing
the residual inflaton density after preheating from a dominant level
down to one where the desired late radiation to matter transition can
be achieved. Moreover, the amount of thermal inflation needed to
achieve this is pretty much the amount already taken as standard in
the literature, for completely different reasons.

Further, since the thermal inflation scenario comes quite close to
failing the nucleosynthesis constraint, it is clear that less drastic
modifications to early Universe dynamics, such as a protracted period
of matter domination due to temporary domination by some long-lived
massive particle species, would not be sufficient to achieve our
goals. Extra periods of early Universe inflation appear essential.

It is of course not so attractive that we have had to invoke a second
period of inflation, in order to unify the first type of inflaton with
dark matter. But at least the thermal inflaton is more grounded in
conventional particle physics, specifically
supersymmetry. Additionally, even conventional high-scale inflation
models may too need thermal inflation in order to solve extra relic
abundance problems such as the gravitino \cite{gravitino}.

The scenario that we have described is based around the quadratic
potential, but the construction is of course more general and can be
applied to a wide range of inflation models. Indeed, at least within
the context of the string landscape, the quadratic potential is
actually quite unattractive as its form has to hold over field values
many times greater than the (reduced) Planck mass, which is the scale
on which we expect the potential to have features \cite{PKL}. It may
be much more plausible to consider inflation as occurring near a
hilltop \cite{hilltop} between neighbouring minima in the landscape;
we anticipate the calculation going through more or less as in this
paper, but perhaps different in the fine numerical details (for
instance, Eq.~(\ref{eq:main}) depends on the inflaton mass, whose
value depends on the shape of the potential during inflation).

Another reason to consider different potentials is that thermal
inflation significantly reduces the number of inflationary $e$-folds
corresponding to the present horizon, perhaps to 40 rather than the
usual 50 to 60 \cite{LL}. This forces the predictions for the
observables $n$ and $r$ further from the slow-roll limit $n=1$ and
$r=0$, and WMAP5 is starting to exert significant observational
pressure against the quadratic potential for low $e$-folding numbers
\cite{Komatsu2008}. While this is not yet conclusive, it certainly
motivates study of potentials which can produce smaller values of $r$.

\begin{acknowledgments}
A.R.L.\ was supported by STFC, C.P.\  by the Swiss Sunburst Fund and by the Marie Curie programme at IAP, and
L.A.U.-L.\ by CONACYT (46195, 47641, 56946), DINPO and
PROMEP-UGTO-CA-3, and by STFC during a visit to
Sussex. L.A.U.-L. is part of the Instituto Avanzado de Cosmolog\'ia
(IAC) collaboration.
\end{acknowledgments}



\end{document}